\documentclass[12pt,a4paper]{article}
\pdfoutput=1
\usepackage[margin=2cm]{geometry}
\usepackage{comment}
\usepackage{amsmath,amssymb,extarrows,graphicx,subfigure,setspace}
\usepackage{cite}
\usepackage{slashed}
\usepackage{color}
\usepackage{braket}
\usepackage{float}
\usepackage[dvipsnames]{xcolor}
\usepackage{bbm}
\makeatother
\usepackage{pdfpages}
\usepackage{blindtext}
\usepackage{amsfonts}
\usepackage{tensor}
\usepackage{graphicx}
\usepackage{pict2e}
\usepackage{amssymb, amsmath,environ}
\usepackage[
colorlinks=true,
linkcolor=blue,
urlcolor=blue,
filecolor=blue,
citecolor=red,
pdfstartview=FitV,
pdftitle={},
pdfauthor={},
pdfsubject={},
pdfkeywords={},
pdfpagemode=None,
bookmarksopen=true
]{hyperref}
\usepackage{tikz}
\usetikzlibrary{arrows,arrows.meta,intersections, calc,positioning,decorations.pathreplacing,decorations.pathmorphing,shapes}
\usetikzlibrary{patterns}
\usetikzlibrary{decorations.markings}
\usetikzlibrary{knots}
\usepackage{epsfig}

\usepackage{hyperref}

\newcommand{\be}{\begin{equation}}
	\newcommand{\bea}{\begin{eqnarray}}
		\newcommand{\eea}{\end{eqnarray}}
	\newcommand{\ba}{\begin{array}}
		\newcommand{\ea}{\end{array}}
	\newcommand{\ee}{\end{equation}}
\newcommand{\bes}{\begin{equation*}}
	\newcommand{\beas}{\begin{eqnarray*}}
		\newcommand{\eeas}{\end{eqnarray*}}
	\newcommand{\bas}{\begin{array*}}
		\newcommand{\eas}{\end{array*}}
	\newcommand{\ees}{\end{equation*}}

\numberwithin{equation}{section}
\begin{document}
\onehalfspacing
\noindent
	
\begin{titlepage}
\vspace{10mm}
	
\vspace*{20mm}
\begin{center}
{\Large {\bf Symmetry-Resolved Relative Entropy of Random States }\\
}
		
\vspace*{15mm}
\vspace*{1mm}
{Mostafa Ghasemi}
		
\vspace*{1cm}

		{\em \hskip -.1truecm Research Center for High Energy Physics, Department of Physics, Sharif University of Technology,\\
			P.O.Box 11155-9161, Tehran, Iran \vskip 5pt }
			{\it 
				School of Particles and Accelerators, Institute for Research in Fundamental Sciences (IPM)\\
				P.O. Box 19395-5531, Tehran, Iran}\\
\vspace*{0.5cm}
{E-mails: {\tt ghasemi.mg@ipm.ir}}
		
\vspace*{1cm}
\end{center}

\date{\today}

\begin{abstract}

We use large-$N$ diagrammatic techniques to calculate the relative entropy of symmetric random states drawn from the Wishart ensemble. These methods are specifically designed for symmetric sectors, allowing us to determine the relative entropy for random states exhibiting $U(1)$ symmetry. This calculation serves as a measure of distinguishability within the symmetry sectors of random states. Our findings reveal that the symmetry-resolved relative entropy of random pure states displays universal statistical behavior. Furthermore, we derive the symmetry-resolved Page curve.  These results deepen our understanding of the properties of these random states.

\end{abstract}

\end{titlepage} 

\section{Introduction} \label{intro}

Random Matrix Theory (RMT) is a mathematical framework that studies the properties of matrices whose entries are random variables. It is a ubiquitous  topic in  mathematics and quantum physics \cite{Mehta:2004RMT,Eynard:2015aea}. RMT has significant implications across various fields, ranging from statistics, number theory, and combinatorics to characterizing the dynamics of chaotic quantum many-body systems and thermalization \cite{DAlessio:2015qtq}, quantum information \cite{Collins:2015qtqI}, the foundations of statistical mechanics \cite{Popescu:2006gts,Gogolin:2015gts}, and providing  toy models for the black hole information problem \cite{Page:1993wv,Hayden:2007cs}.

 A random quantum state is a state that is sampled according to some random distribution, typically the uniform (Haar) distribution on the space of pure states  or the space of density matrices for mixed states.
One of the main structures studied in the realm of random states is the quantum entanglement structure. The entanglement structure of random states was initially explored in the work of Page \cite{Page:1993df}, where the average von Neumann entanglement entropy of random pure states distributed according to the Haar measure was considered. 
 For the bipartite system $A\cup B$, and a large Hilbert space $\mathcal{H}= \mathcal{H}_{A}\otimes \mathcal{H}_{B}$, Page's formula gives
\begin{align}
	\mathcal{S}\approx\text{min}\left( \log d_{A},\log d_{B}\right)+\cdots .
\end{align}
Here, $d_{A}$ and $ d_{B}$ represent the dimensions of $\mathcal{H}_{A}$ and $\mathcal{H}_{B}$, respectively. The average entanglement entropy of a subsystem, up to subleading corrections, is the maximal one. This implies that the bipartite entanglement entropy of typical states is nearly maximal. The average  entanglement entropy, when plotted against the subsystem’s size, forms a Page curve. 
The Page curve describes how the entanglement entropy of Hawking radiation evolves over time as a black hole evaporates. Initially, it increases but eventually decreases as the black hole loses information, providing insights into how information might be recovered from black holes. Page’s
formula provides a qualitative description for the behavior of  typical Hamiltonians or
states generated by sufficiently chaotic dynamics. It is believed that these systems can be well described by random states. For a detailed review, refer to \cite{DAlessio:2015qtq, Bianchi:2021aui}.

Another quantum information- theoretic quantity that can be useful in this context is the relative entropy. Relative entropy is a measure of distinguishability for quantum states and is essential  in quantum information theory \cite{Vedral:2002zz} and quantum statistical mechanics \cite{Wehrl:1978zz}. Relative entropy has favorable properties, like monotonicity and positivity. 
These properties, make the relative entropy a valuable
 tool in various areas of physics, including quantum field theory \cite{Faulkner:2016mzt, Balakrishnan:2017bjg,Casini:2016udt, Casini:2019qst}, conformal field theory \cite{Lashkari:2014yva,Lashkari:2015dia, 	Ruggiero:2016khg, Sarosi:2016oks}, boundary conformal field theory \cite{Ghasemi:2024wcq} holography \cite{Blanco:2013joa, Jafferis:2014lza,Jafferis:2015del,Lashkari:2016idm}, quantum gravity \cite{Casini:2008cr,Wall:2011hj,Bousso:2014sda,Bousso:2014uxa,Longo:2018zib}, and  random states \cite{Kudler-Flam:2021rpr,Kudler-Flam:2021alo}.

The relative entropy  between two density matrices $\rho$ and $\sigma$ is defined as
\begin{align}
D(\rho \lvert \rvert \sigma ) := \mathrm{Tr} \left[ \rho\left( \log \rho - \log \sigma\right) \right].
\end{align}
Here, $ D(\rho \lvert \rvert \sigma ) = 0$ if and only if $\rho = \sigma$. Importantly, the relative entropy is monotonic under quantum operations 
\begin{align}
D(\mathcal{N}(\rho) \lvert \rvert \mathcal{N}(\sigma) ) \leq D(\rho \lvert \rvert \sigma ),
\end{align}
where $\mathcal{N}$ represents any completely-positive trace-preserving map. A significant quantum operation is the partial trace, and monotonicity in this context implies that density matrices become less distinguishable as  more information about them is discarded. In this context, the author \cite{Kudler-Flam:2021rpr} calculate the relative entropy for Haar random states using a large-N diagrammatic technique,
\begin{align}
	&D(\rho_A \lvert \rvert \sigma_A) = 1+\frac{d_A}{2 d_B}+\left(\frac{d_B}{d_A}-1\right) \log
	\left(1-\frac{d_A}{d_B}\right).
	\label{maineq}
\end{align}
The relative entropy monotonically
increases with $d_A/d_B$ and reaches a value of  $3/2$
when $d_A=d_B$. This behavior demonstrates
the monotonicity of relative entropy under the partial trace as expected.

Quantum chaos and thermalization are fascinating concepts in quantum physics that relate to distinguishability, randomness in states, and the behavior of quantum systems as they evolve over time.
Random states are typical states in high-dimensional Hilbert spaces and play a crucial role in understanding the behavior of thermalized quantum systems. As quantum systems evolve chaotically, their states become increasingly random, resembling thermal states.
 Relative entropy is useful in tracking the loss of information in quantum chaotic systems. As a chaotic system evolves, information about its initial state becomes scrambled and lost, making the state more random and more similar to a thermal state.
 Random states, being effectively indistinguishable from thermal states, have low relative entropy with respect to the thermal state. In the presence of symmetry, it is interesting to explore the distinguishability within each symmetry sector. In the next section, we will address the interplay between symmetry and entanglement.

\subsection{Symmetry-Resolved Relative Entropy}
Entanglement and symmetry are two pillars of modern physics. 
One interesting problem in this context involves understanding the relationship between entanglement and symmetry. The presence of symmetry can impact the distribution of entanglement. It's important to explore how entanglement is distributed within the symmetry sectors of a theory. One of the main tools used to study this is the symmetry-resolved entanglement entropy \cite{Goldstein:2017bua}, which determines the contribution of each symmetry sector associated with a given global symmetry to the total entanglement entropy.

More generally, in the presence of a global symmetry, the density matrix $\rho$ of the total system commutes with a conserved charge $\hat{Q}$, $[\rho ,\hat{Q}]=0$. By
tracing over the degrees of freedom of the subsystem $B$, we find that $[\rho_{A} ,\hat{Q}_{A}]=0$, where $\hat{Q}_{A}$ is the total charge in the $A$ subsystem. This means that  the reduced density matrix $\rho_{A}$ can be decomposed into block diagonal form associated with each charge sector, 
\begin{equation}
\label{SRDE}
\rho_{A}=\bigoplus_{Q} P(Q_{A})\rho_{A}(Q_{A}),
\end{equation}
and the charged partition function becomes
\begin{equation}
\mathcal{Z}_n(Q_A)=\mathrm{Tr}{\rho^n_A \mathcal{P}_{Q_A}}=\int_{-\pi}^{\pi}{\frac{\textrm{d}\alpha}{2\pi} {\mathcal{Z}}_n(\alpha)e^{-i \alpha Q_{A}}}.
\end{equation}
$\mathcal{P}_{Q_A}$ denotes the projection into the subspace of states of region $A$ with charge $\hat{Q}_A$. Symmetry-resolved R\'{e}nyi entropies are defined as, 
\begin{equation}
S_{n}(Q_A) = \frac{1}{1-n}\log \left[ \frac{\mathcal{Z}_n(Q_A)}{\mathcal{Z}_1^{n}(Q_A)} \right],
\end{equation}
and symmetry-resolved entanglement entropy is obtained as,
\begin{equation}
S(Q_A)=\lim_{n\rightarrow1} S_{n}(Q_A).
\end{equation}	
Probability is then recovered as $P(Q_A)=\mathcal{Z}_1(Q_A)$.
Regarding the decomposition of the reduced density matrix (\ref{SRDE}), the total entanglement entropy can be divided into two terms,
\begin{eqnarray}
	\label{SYM-RES1}
	S_{E} &=& \sum_{Q_A}P(Q_A)S(Q_A)-\sum_{Q_A}P(Q_A)\log (P(Q_A)) \nonumber \\
	&&
	=S^{c}+S^{n} . \ 
\end{eqnarray}
$S^{c}$  represents the configurational entropy \cite{lukin:2018}, which is defined as the weighted sum of the entropies in each sector.
 On the other hand, $S^{n}$, known as the number entropy \cite{lukin:2018,wv-03,kusf2,bhd-18}, quantifies the entropy caused by the fluctuations of charge within subsystem $A$. 

Over the past few years, this subject has been the focus of intense research activity, including studies on topics such as 1+1-dimensional conformal field theories  \cite{Goldstein:2017bua, Xavier:2018kqb, Capizzi:2020jed,Calabrese:2021wvi,Ghasemi:2022jxg,Ares:2022gjb,DiGiulio:2022jjd,Kusuki:2023bsp,Gaur:2023yru}. Additionally see   \cite{Laflorencie2014,Belin:2013uta,Zhao:2020qmn,Weisenberger:2021eby,Zhao:2022wnp,DiGiulio:2023nvz,Milekhin:2021lmq, Hejazi:2021yhz,Horvath:2023xoh,Fossati:2023zyz,Feldman:2024qif} for further references.
One main feature emerging from the literature is that, due to conformal invariance, the entanglement entropy is equally distributed among different sectors. Strictly speaking, at the leading order, the symmetry-resolved entanglement entropy is independent of the charge sector.

Similar to symmetry-resolved entanglement entropy, the concept of symmetry-resolved relative entropy (SRRE) can be defined \cite{Chen:2021pls,Capizzi:2021zga}. Symmetry-resolved relative entropy provides measures of the distinguishability of two states within the same symmetry sector. For two reduced density matrices, denoted as $\rho$ and $\sigma$, the symmetry-resolved relative entropy corresponds to each sector $Q$ defined as
\begin{align}\label{SR-RE}
D(\rho(Q) \lvert \rvert \sigma(Q)) := \mathrm{Tr} \left[ \rho(Q)\left( \log \rho(Q) - \log \sigma(Q)\right) \right].
\end{align}
The symmetry-resolved relative entropies satisfy the sum rule,
\begin{align}\label{SR-RE2}
D(\rho \lvert \rvert \sigma)= \sum_{Q}P^{\rho}(Q)D(\rho(Q) \lvert \rvert \sigma(Q))+\sum_{Q}P^{\rho}(Q)\log\frac{P^{\rho}(Q)}{P^{\sigma}(Q)}
\end{align}
where $P^{\rho}(Q)\equiv \text{tr}\left(\rho \Pi_{Q}\right)$ and $P^{\sigma}(Q)\equiv \text{tr}\left(\sigma \Pi_{Q}\right)$. From a replica perspective, the symmetry-resolved relative entropy can be obtained as the replica limit $n \rightarrow 1$ of the $n$-th R\'enyi entropy of the sector $Q$,
\begin{align}\label{SR-REE1}
D(\rho(Q) \lvert \rvert \sigma(Q))= -\partial_{n}\,  \log \left(\frac{\mathrm{Tr}\left( \rho(Q) \sigma^{n-1}(Q)\right) }{\text{tr}\rho^{n}(Q)} \right) \Big|_{n \to 1}.
\end{align}

The interesting question that motivated this work is: What is the thermalization process in individual system charge sectors? 
Quantum thermalization in isolated many-body systems with global symmetry, and the charged black hole information problem, motivate us to study the relative entropy of symmetric random quantum states. A more fundamental quantity in this context is the symmetry-resolved relative entropy. It not only encodes information about the interplay between symmetry and entanglement, but also provides the statistical distance between two states in a given symmetry sector. In this work, we investigate the symmetry-resolved relative  entropy in the 
Haar-random states
with a U(1) symmetry.
Since $X$ is a Gaussian random variable, all of its moments can be decomposed in terms of its second moment.  Using the large-$N$ diagrammatic techniques, we derive explicit analytic formulas for the corresponding symmetry-resolved
Page curves and symmetry-resolved relative entropy. This paper is organized as follows: in section \ref{Sec-2}, we introduce the $U(1)$ symmetric random states and evaluate the symmetry-resolved Page curve. In section \ref{Sec-3}, we evaluate the symmetry-resolved relative entropy for the $U(1)$ symmetric random states.
A summary of the results and further directions for research are included in Section \ref{Sec-4}.

\section{Symmetric Random Pure States.}\label{Sec-2}
 
In this section, we will study an ensemble of Haar random pure states with a $U(1)$ charge and generator $\hat{Q}$ based on the diagrammatic approach \cite{Kudler-Flam:2021alo, Kudler-Flam:2021rpr,Shapourian:2020mkc}. Given a global symmetry, the Hilbert space decomposes as 
\begin{align}
\mathcal{H}=\bigoplus_{Q=0}\mathcal{H}(Q),
\end{align}
where $\mathcal{H}(Q)$ denotes the eigenspace of the charge operator $\hat{Q}$ corresponding to eigenvalue $Q$.
 The total charge  $Q$  of the whole system has an additive property and can be written as the sum of the charges $q$ of its constituent subsystems. Therefore, for the bipartite system, $A\bar{A}$, a projector with a given charge $Q$ for the whole system can be written in terms of the projectors of its constituents $A$ and $\bar{A}$ subsystems in the following way:
\begin{align}\label{eq:additivity}
	\hat \Pi^{(Q)}_{A\bar{A}}= \sum_{
		q_A+\bar{q}_{\bar{A}}=Q } \hat \Pi^{(q_A)}_{A}\otimes \hat \Pi^{(\bar{q})}_{\bar{A}}\ . 
\end{align}
Consequently, the Hilbert space  $\mathcal{H}(Q)$ has a decomposition as
\begin{align}
\mathcal{H}(Q)=\bigoplus_{q=0}^{Q}\mathcal{H}_{A}(q)\otimes \mathcal{H}_{\bar{A}}(\bar{q}),
\end{align}
where $\mathcal{H}_{A}(q)$ represents the eigenspace of the charge operator $\hat{Q}_{A}$ with an eigenvalue $q$, and $\mathcal{H}_{\bar{A}}(\bar{q})$ denotes the eigenspace of the charge operator $\hat{Q}_{\bar{A}}$ with an eigenvalue $\bar{q}=Q-q$. In the upcoming discussion, we will utilize the random matrix formalism and large-$N$ diagrammatic techniques \cite{Kudler-Flam:2021alo, Kudler-Flam:2021rpr,Shapourian:2020mkc} adapted for symmetric random states, to derive symmetry-resolved Page curves and symmetry-resolved relative entropy for this class of random states.

Let's begin with a symmetric Haar random pure state $\ket{\Psi^{(Q)}}$ with a definite charge $Q$ on a bipartite Hilbert space $\mathcal{H}(Q)$, i.e.,  $\hat \Pi^{(Q^{\prime})}\ket{\Psi^{(Q)}}= \delta_{Q^{\prime},Q} \ket{\Psi^{(Q)}}$,
\begin{align}
\label{}
\ket{\Psi^{(Q)}  } = \sum_{\substack{i,\alpha \\
		q + \bar{q} = Q}}^{d_A(q),d_{\bar{A}}(\bar{q})} \, X_{i_{q}, \alpha_{\bar{q}}}^{(q, \bar{q})}\ket{i_{q}}_A \otimes \ket{\alpha_{\bar{q}}}_{\bar{A}},
\end{align}
where $\left\lbrace\ket{i_{q}}_A  \right\rbrace $ and $\left\lbrace \ket{\alpha_{\bar{q}}}_{\bar{A}} \right\rbrace $ are orthonormal bases for the sub-Hilbert spaces $\mathcal{H}_{A}(q)$ and $\mathcal{H}_{\bar{A}}(\bar{q})$ with dimensions $d_A(q)$ and $d_{\bar{A}}(\bar{q})$, respectively.   We will assume that these sub-Hilbert spaces are independently large.  The dimension of $\mathcal{H}(Q)$ is $d(Q)\equiv\text{dim}\mathcal{H}(Q)=\sum_{q} d_A(q) d_{\bar{A}}(\bar{q})$.

The coefficients $ X_{i_{q}, \alpha_{\bar{q}}}^{(q, \bar{q})}$'s are independent complex Gaussian random variables within the symmetry sector. Their joint probability distribution is defined as\footnote{In general, we define the ensemble of random states $\ket{\Psi^{(Q)}}$ drawn out of the uniform Haar distribution over the set of all states in $\mathcal{H}(Q)$. Within this subspace, we randomly select  random pure states using a distribution of complex Gaussian random variables.} 
\begin{align}
P(\{ X_{i_{q}, \alpha_{\bar{q}}}^{(q, \bar{q})}\}) = \mathcal{Z}(Q)^{-1}\exp\left[-d_A(q)d_B(\bar{q}) \mathrm{Tr} \left(X^{(q,\bar{q})} X^{(q, \bar{q})\dagger}\right)\right],
\end{align}
where $\mathcal{Z}(Q)$ is the normalization factor. 
$X^{(q, \bar{q})}$ represents the rectangular matrix whose elements  are $ X_{i_{q}, \alpha_{\bar{q}}}^{(q,\bar{q})}$ in the basis $\left\lbrace\ket{i_{q}}_A  \right\rbrace $ and $\left\lbrace \ket{\alpha_{\bar{q}}}_{\bar{A}} \right\rbrace $. 

The additivity of the symmetry charge implies that any reduced density matrix \(\hat{\rho}_A\) obtained from a symmetric pure state \(\ket{\Psi^{(Q)}}\) through partial tracing, \(\hat{\rho}_A = \mathrm{Tr}_{\bar{A}} \ket{\Psi^{(Q)}}\bra{\Psi^{(Q)}}\), is also symmetric and can be expressed as shown in equation (\ref{SRDE}). The random reduced density matrix on \(\mathcal{H}_A(q)\) is therefore given by
\begin{align}
\rho_A (q)= \frac{X^{(q,\bar{q})} X^{(q,\bar{q})\dagger}}{\mathrm{Tr}(X^{(q,\bar{q})} X^{(q, \bar{q})\dagger})}.
\end{align}
In the limit of large Hilbert space dimension, the denominator is sharply peaked around unity, $\mathrm{Tr}(X^{(q,\bar{q})} X^{(q, \bar{q})\dagger})=1+\delta$. Therefore, to the leading order in $(1/d_A(q)d_{\bar{A}(\bar{q})})$, we can write 
\begin{align}
\rho_A (q)\simeq X^{(q,\bar{q})} X^{(q,\bar{q})\dagger},
\end{align}
which defines the Wishart ensemble on each symmetry sector. In a diagrammatic language for random states 
 the representation of the matrix element of the pure state density matrix $\rho=\ket{\Psi^{(Q)}}\bra{\Psi^{(Q)}}$ associated with the symmetry sector is 
\begin{align}
\left[\rho\right]_{i_{q}\alpha_{\bar{q}},j_{q}\beta_{\bar{q}}}=& \sum_{\substack{i,j,\alpha,\beta \\
		q + \bar{q} = Q, q^{\prime} + \bar{q}^{\prime} = Q}}^{d_A(q),d_{\bar{A}}(\bar{q})} \, X_{i_{q}, \alpha_{\bar{q}}}^{(q, \bar{q})}X_{j_{q^{\prime}}, \beta_{\bar{q}^{\prime}}}^{(q, \bar{q})\ast}
\nonumber \\ 
=&
\sum_{\substack{i,j,\alpha,\beta\\
		q + \bar{q} = Q, q^{\prime} + \bar{q}^{\prime} = Q}}^{d_A(q),d_{\bar{A}}(\bar{q})} \,
\tikz[scale=0.8,baseline=-0.5ex]{
	\draw[dashed] (0.,0.2)--  (0,-0.2);
	\draw[dashed]  (2,0.2) -- (2,-0.2);
	\draw (-0.3,0.3)-- (-0.3,-0.25);
	\draw (2.3,-0.25)-- (2.3,0.3);
	\draw (-0.3,0.5) circle (0.) node[anchor=center] {\footnotesize$i_{q}$};
	\draw (2.5,0.5) circle (0.) node[anchor=center] {\footnotesize$j_{q^{\prime}}$};
	\draw (0.1,0.5) circle (0.) node[anchor=center] {\footnotesize$ \alpha_{\bar{q}} $};
	\draw (2.0,0.5) circle (0.) node[anchor=center] {\footnotesize$ \beta_{\bar{q}^{\prime}} $};
}\ .
\end{align}
The left and right pairs of lines represent the bra and ket states, respectively. The solid and dashed lines correspond to subsystems $A$ and ${\bar{A}}$, respectively. 
The reduced density matrix in the sector given by $q$ is obtained by taking the partial trace over $\mathcal{H}_{\bar{A}}( \bar{q})$ 
\begin{equation}\label{RED_A_expansion}
\begin{aligned}
\rho_{A}(q)  &= \sum_{i_{q}, j_{q}} \sum_{\alpha_{\bar{q}}} 
X_{i_{q}, \alpha_{\bar{q}}}^{(q, \bar{q})}\; X_{j_{q}, \alpha_{\bar{q}}}^{(q, \bar{q})\ast} \ \left| i_{q} \right\rangle \left\langle  j_{q} \right|,
\end{aligned}
\end{equation}
with $\bar{q} = Q - q$. In diagrammatic formalism, by definition,
matrix manipulations are performed at the bottom edge of the diagram, while ensemble averaging is done at the top of the diagram with  appropriate propagators carrying weight. For instance, the matrix elements of the density matrix $ \rho_{A}(q)$ are
\begin{equation}
\left[ \rho_{A}(q)\right]_{i_{q}, j_{q}} =\sum_{\alpha_{\bar{q}=1}}^{d_B(\bar{q})} 
X_{i_{q}, \alpha_{\bar{q}}}^{(q, \bar{q})}\; X_{j_{q}, \alpha_{\bar{q}}}^{(q, \bar{q})\ast}
:=
\,
\tikz[scale=0.8,baseline=-0.5ex]{
	\draw[dashed] (0.,0.2)-- (0,0);
	\draw[dashed] (0,0) -- (2,0);
	\draw[dashed]  (2,0.2) -- (2,0);
	\draw (-0.3,0.2)-- (-0.3,-0.15);
	\draw (2.3,-0.15)-- (2.3,0.2);
	\draw (-0.5,0.5) circle (0.) node[anchor=center] {\footnotesize$i_{q}$};
	\draw (2.5,0.5) circle (0.) node[anchor=center] {\footnotesize$j_{q}$};
	\draw (0.2,0.5) circle (0.) node[anchor=center] {\footnotesize$ \alpha_{\bar{q}} $};
	\draw (1.8,0.5) circle (0.) node[anchor=center] {\footnotesize$ \alpha_{\bar{q}} $};
}\ .
\end{equation}
and  the propagators are represented as 
\begin{align}
\,
\tikz[baseline=0ex]{
	\draw[dashed] (1.0,0) arc (0:180:0.5);
	\draw (1.2,0.) arc (0:180:0.7);
		\draw (1.2,0.4) circle (0.) node[anchor=center] {\footnotesize$q$};
		\draw (-0.1,0.10) circle (0.) node[anchor=center] {\footnotesize$\bar{q}$};
}\
:= \braket{X_{i_{q}, \alpha_{\bar{q}}}^{(q, \bar{q})}X_{j_{q^{\prime}}, \beta_{\bar{q}^{\prime}}}^{(q^{\prime}, \bar{q}^{\prime})\ast}}=  \frac{\delta_{qq^{\prime}}\delta_{\bar{q}\bar{q}^{\prime}}\delta_{i_{q}j_{q^{\prime}}} \delta_{\alpha_{\bar{q}}\beta_{\bar{q}^{\prime}}}}{d_A(q) d_{\bar{A}}(\bar{q})}. 
\label{eq:doubleline1}
\end{align}
These renormalization factors ensure that $\rho_{A}(q)$ has a unit trace on average. Then, take the trace of the density matrix $\rho_{A}(q)$ as follows:
\begin{align}
\braket{\mathrm{Tr}\rho_A(q)}=
\,
\tikz[baseline=0ex]{
	\draw[dashed] (1.0,0) arc (0:180:0.5);
	\draw[dashed] (0,0) -- (1,0);
	\draw (1.2,0.) arc (0:180:0.7);
	\draw (-0.2,0.0)-- (-0.2,-0.15)--(1.2,-0.15)-- (1.2,0.0);
	\draw (1.2,0.4) circle (0.) node[anchor=center] {\footnotesize$q$};
	\draw (-0.1,0.10) circle (0.) node[anchor=center] {\footnotesize$\bar{q}$}
} \  = d_{A}(q)\, d_{\bar{A}}(\bar{q}) \ \frac1{d_{A}(q) \,d_{\bar{A}}(\bar{q})} = 1,
\end{align}
The quantum numbers for each symmetry sector are explicitly displayed on the lines. Each closed loop, whether solid or dashed, contributes a factor equal to the dimension of the corresponding sector. It's important to note that, according to the diagrammatic rules for averaging, we must sum over all possible contractions of the bras and kets. Furthermore, for every insertion of the density matrix, we incorporate a factor of $(d_{A}(q)d_{\bar{A}}(\bar{q}))^{-1}$.

\subsection{Symmetry-resolved Page's curve}\label{}

To compute the ensemble average of the symmetry-resolved relative entropy, we first need to compute the ensemble average of the symmetry-resolved R\'enyi entropy. 
This can be done by representing the symmetry-resolved R\'enyi entropy diagrammatically as follows
\begin{align}
 	\label{renyi_diag}
 	\text{tr} \left[\rho_{A}^{n} \right]=
 	\tikz[scale=0.75,baseline=-0.5ex]{
 		\draw[dashed] (0,0) -- (1,0);
 		\draw (0,0)-- (0,.15);
 		\draw (1,0)-- (1,.15);
 		\draw (-0.2,0.15)-- (-0.2,-0.35);
 		\draw (1.2,-0.15)-- (1.2,0.15);
 		\draw[dashed,black] (2,0) -- (3,0);
 		\draw[black] (2,0)-- (2,0.15);
 		\draw[black] (3,0)-- (3,0.15);
 		\draw[black] (1.8,-.15)-- (1.8,0.15);
 		\draw[black] (3.2,-0.15)-- (3.2,0.15);
 		\draw[dashed,black] (5,0) -- (6,0);
 		\draw[black] (5,0)-- (5,0.15);
 		\draw[black] (6,0)-- (6,0.15);
 		\draw[black] (4.8,0.15)-- (4.8,-0.15);
 		\draw[black] (6.2,-0.35)-- (6.2,0.15);
 		\draw (1.2,-0.15)--(1.8,-0.15);
 		\draw[black] (3.2,-0.15)--(3.45,-0.15);
 		\draw (-0.2,-0.35)-- (6.2,-0.35);
 		\draw[black] (4.5,-0.15)--(4.8,-0.15);
 		\node[] at (4.,0.2) {$\cdots$};
 	}
 	\, .
 \end{align}
  Ensemble averaging involves summing all possible contractions of the same type of bra and kets with each other. We assume that the dimensions of the sub-Hilbert spaces are large, with \(d_A, d_B \propto N\rightarrow \infty\), but their relative sizes, \(d_A/d_B\), are finite. In this limit, the dominant leading diagrams that maximize the number of loops are planar diagrams. In the double-line notation, these diagrams correspond to the standard large-N topological expansion. In the context of enumerative combinatorics and probability theory,
  planar diagrams correspond to non-crossing permutations. As an example, for $n = 2$, we have
\begin{align}
\label{eq:tr_r2}
\mathrm{Tr}(\rho_A(q))^2= 
\,
\tikz[baseline=0ex]{
	\draw[dashed] (0,0.2)-- (0,0);
	\draw[dashed] (0,0) -- (1,0);
	\draw[dashed]  (1,0.2) -- (1,0);
	\draw (-0.2,0.2)-- (-0.2,-0.25);
	\draw (1.2,-0.1)-- (1.2,0.2);
	\draw (-0.3,0.5) circle (0.) node[anchor=center] {\footnotesize$i_{q}$};
	\draw (1.3,0.5) circle (0.) node[anchor=center] {\footnotesize$i_{q}$};
	\draw (0.1,0.5) circle (0.) node[anchor=center] {\footnotesize$ \alpha_{\bar{q}} $};
	\draw (0.9,0.5) circle (0.) node[anchor=center] {\footnotesize$ \alpha_{\bar{q}} $};
	\draw[dashed] (2,0.2)-- (2,0);
	\draw[dashed] (2,0) -- (3,0);
	\draw[dashed]  (3,0.2) -- (3,0);
	\draw (1.8,0.2)-- (1.8,-0.1);
	\draw (3.2,-0.25)-- (3.2,0.2);
	\draw (1.7,0.5) circle (0.) node[anchor=center] {\footnotesize$i_{q}$};
	\draw (3.3,0.5) circle (0.) node[anchor=center] {\footnotesize$i_{q}$};
	\draw (2.1,0.5) circle (0.) node[anchor=center] {\footnotesize$ \alpha_{\bar{q}} $};
	\draw (2.9,0.5) circle (0.) node[anchor=center] {\footnotesize$ \alpha_{\bar{q}} $};
	\draw (1.2,-0.1)-- (1.8,-0.1);
	\draw (-0.2,-0.25) -- (3.2,-0.25);
}\ .
\end{align}
The ensemble average is derived by a sum of the two possible contractions
\begin{align}
\label{purity_diagram}
\overline{\mathrm{Tr}\rho_A(q)^2} &=
\,
\tikz[scale=0.8,baseline=0.5ex]{
	\draw[dashed] (0,0) -- (1,0);
	\draw (-0.2,0.)-- (-0.2,-0.25);
		\draw (0.6,0.2) circle (0.) node[anchor=center] {\footnotesize$ \bar{q} $};
	\draw (1.2,-0.1)-- (1.2,0.); 
		\draw (1.7,0.5) circle (0.) node[anchor=center] {\footnotesize$q$};
	\draw[dashed] (2,0) -- (3,0);
	\draw (1.8,0.)-- (1.8,-0.1);
	\draw (3.2,-0.25)-- (3.2,0.);
	\draw (1.2,-0.1)-- (1.8,-0.1);
	\draw (-0.2,-0.25) -- (3.2,-0.25);
	\draw[dashed] (1.0,0) arc (0:180:0.5);
	\draw[dashed] (3.0,0) arc (0:180:0.5);
	\draw (1.2,0) arc (0:180:0.7);
	\draw (3.2,0) arc (0:180:0.7);
}
\ \,
+
\ \,
\tikz[scale=0.7,baseline=0.5ex]{
	\draw[dashed] (0,0) -- (1,0);
	\draw (-0.2,0.)-- (-0.2,-0.25);
	\draw (1.2,-0.1)-- (1.2,0.);
	\draw (0.6,0.2) circle (0.) node[anchor=center] {\footnotesize$ \bar{q} $};
	\draw[dashed] (2,0) -- (3,0);
	\draw (1.8,0.)-- (1.8,-0.1);
	\draw (3.2,-0.25)-- (3.2,0.);
	\draw (1.2,-0.1)-- (1.8,-0.1);
	\draw (-0.2,-0.25) -- (3.2,-0.25);
	\draw[dashed] (2.0,0) arc (0:180:0.5);
	\draw[dashed] (3.0,0) arc (0:180:1.5);
	\draw (1.8,0) arc (0:180:0.3);
	\draw (3.2,0) arc (0:180:1.7) (3.4,0.2) circle (0.) node[anchor=center] {\footnotesize$ q $};
},
\ \,
\nonumber
\\
&=\frac{1}{d_{A}(q)} + \frac{1}{d_{\bar{A}}(\bar{q})},
\end{align}
Similarly, for $n = 3$, we found that
\begin{align}
\label{eq:tr_r3}
\overline{ \mathrm{Tr} \left( \rho_{A}(q) \right)^3 } 
=&
\
\tikz[scale=0.5,baseline=-0.5ex]{
	\draw[dashed] (0,0) -- (1,0)(2.4,0.1) circle (0.) node[anchor=center] {\footnotesize$ \bar{q} $};
	\draw (-0.2,0.)-- (-0.2,-0.35);
	\draw (1.2,-0.15)-- (1.2,0.)(3.4,0.7) circle (0.) node[anchor=center] {\footnotesize$ q $};
	\draw[dashed] (2,0) -- (3,0);
	\draw (1.8,0.)-- (1.8,-0.15);
	\draw (3.2,-0.15)-- (3.2,0.);
	\draw[dashed] (4,0) -- (5,0);
	\draw (3.8,0.)-- (3.8,-0.15);
	\draw (5.2,-0.35)-- (5.2,0.);
	\draw (1.2,-0.15)--(1.8,-0.15);
	\draw (3.2,-0.15)--(3.8,-0.15);
	\draw (-0.2,-0.35)-- (5.2,-0.35);
	\draw[dashed] (1.0,0) arc (0:180:0.5);
	\draw[dashed] (3.0,0) arc (0:180:0.5);
	\draw[dashed] (5.0,0) arc (0:180:0.5);
	\draw (1.2,0) arc (0:180:0.7);
	\draw (3.2,0) arc (0:180:0.7);
	\draw (5.2,0) arc (0:180:0.7);
} \,
+
3\times\,
\tikz[scale=0.4,baseline=0.5ex]{
	\draw[dashed] (0,0) -- (1,0);
	\draw (-0.2,0.)-- (-0.2,-0.35);
	\draw (1.2,-0.15)-- (1.2,0.)(3.4,1.1) circle (0.) node[anchor=center] {\footnotesize$ q $};
	\draw[dashed] (2,0) -- (3,0);
	\draw (1.8,0.)-- (1.8,-0.15);
	\draw (3.2,-0.15)-- (3.2,0.)(2.4,0.5) circle (0.) node[anchor=center] {\footnotesize$ \bar{q} $};
	\draw[dashed] (4,0) -- (5,0);
	\draw (3.8,0.)-- (3.8,-0.15);
	\draw (5.2,-0.35)-- (5.2,0.);
	\draw (1.2,-0.15)--(1.8,-0.15);
	\draw (3.2,-0.15)--(3.8,-0.15);
	\draw (-0.2,-0.35)-- (5.2,-0.35);
	\draw[dashed] (2.0,0) arc (0:180:0.5);
	\draw[dashed] (3.0,0) arc (0:180:1.5);
	\draw[dashed] (5.0,0) arc (0:180:0.5);
	\draw (1.8,0) arc (0:180:0.3);
	\draw (3.2,0) arc (0:180:1.7);
	\draw (5.2,0) arc (0:180:0.7);
}  
\nonumber \\ 
& +
\,
\tikz[scale=0.4,baseline=0.5ex]{
	\draw[dashed] (0,0) -- (1,0);
	\draw (-0.2,0.)-- (-0.2,-0.35);
	\draw (1.2,-0.15)-- (1.2,0.)(3.4,1.8) circle (0.) node[anchor=center] {\footnotesize$ \bar{q} $};
	\draw[dashed] (2,0) -- (3,0);
	\draw (1.8,0.)-- (1.8,-0.15);
	\draw (3.2,-0.15)-- (3.2,0.);
	\draw[dashed] (4,0) -- (5,0);
	\draw (3.8,0.)-- (3.8,-0.15);
	\draw (5.2,-0.35)-- (5.2,0.);
	\draw (1.2,-0.15)--(1.8,-0.15);
	\draw (3.2,-0.15)--(3.8,-0.15);
	\draw (-0.2,-0.35)-- (5.2,-0.35);
	\draw[dashed] (4.0,0) arc (0:180:0.5);
	\draw[dashed] (2.0,0) arc (0:180:0.5);
	\draw[dashed] (5.0,0) arc (0:180:2.5);
	\draw (3.8,0) arc (0:180:0.3);
	\draw (1.8,0) arc (0:180:0.3);
	\draw (5.2,0) arc (0:180:2.7)(5.5,0.2) circle (0.) node[anchor=center] {\footnotesize$ q $};
}  \,
\nonumber \\ 
=& 
\frac{1}{ d_{A}(q)} 
+ \frac{3}{d_{A}(q) d_{\bar{A}}(\bar{q})}
+ \frac{1}{d_{\bar{A}}(\bar{q})}.
\end{align}
Similarly, the higher moments can be computed by contracting various  bras and kets of the same type with each other. In general, the sum runs over all possible contractions, 
so the moments can be expressed as a sum over the permutation group
\begin{align}
\overline{\mathrm{Tr} \left[\rho_{A}^{{n}} \right]} = \frac{1}{(d_A(q) d_{\bar{A}}(\bar{q}))^n} \sum_{\tau \in S_n} (d_A(q))^{C(\eta^{-1} \circ \tau)}(d_{\bar{A}}(\bar{q}))^{C(\tau)},
\label{renyi_sum}
\end{align}
where $C(\cdot)$ represents the number of cycles in the given permutation and $\eta$ denotes the cyclic permutation. Each permutation corresponds to a diagram, with the cycle structure determining which bra is contracted with which ket.
When the sub-Hilbert spaces are large, only the contribution of the non-crossing permutations is dominant. In other words, these are the transformations that maximize $C(\eta^{-1} \circ \tau) + C( \tau)$. For non-crossing permutations, we have $C(\eta^{-1} \circ \tau) + C( \tau) = n+1$.  The number of such permutations with $C(\eta^{-1} \circ \tau) = k$ is given by the Narayana number \cite{KREWERAS1972333, SIMION2000367}
\begin{align}
N_{n,k} = \frac{1}{n}\binom{n}{k}\binom{n}{k-1}.
\end{align}
Thus, the sum  (\ref{renyi_sum}) can be re-expressed as
\begin{align}
\overline{ \mathrm{Tr} (\rho_A(q))^n } &= \frac{1}{(d_A(q) d_{\bar{A}}(\bar{q}))^n} \sum_{k = 1}^n N_{n,k} (d_A(q))^{k}(d_{\bar{A}}(\bar{q}))^{n+1-k}
\nonumber
\\
&= (d_A(q))^{1-n} \, _2F_1\left(1-n,-n;2;\frac{d_A(q)}{d_{\bar{A}}(\bar{q})}\right),
\label{renyi_hyper}
\end{align}
where $\, _2F_1$ is a hypergeometric function. This expression gives the \textit{the symmetry-resolved Page's curve} \cite{Bianchi:2019stn,Murciano:2022lsw,Lau:2022hvc}.
\begin{align}
\overline{ S_{A}(q)}
&= \log d_A(q)-\frac{ d_A(q)}{2d_{\bar{A}}(\bar{q})},\, 
\label{SR-Page}
\end{align}
This expression characterizes the ensemble average bipartite symmetry-resolved entanglement entropies for ensembles of Haar-random pure states with a $U(1)$-symmetry. This result can be interpreted in the context of the information paradox in charged black holes or thermalization in each symmetry sector.

\section{Symmetry-resolved Relative Entropy of Random States}\label{Sec-3}
By using a replica trick (\ref{SR-REE1}), we can compute the symmetry-resolved relative entropy. In the case of relative entropy, we have two distinct symmetry-resolved density matrices, $\rho_A(q)$ and $\sigma_A(q)$, associated with the symmetry sector $q$. These matrices must be averaged separately within the ensemble. To avoid confusion, we use black and red colors to represent $\rho_A(q)$ and $\sigma_A(q)$, respectively.

To calculate the symmetry-resolved relative entropy, we need to determine the diagram that represents the overlap between symmetry-resolved density matrices raised to arbitrary powers. For example, when considering two density matrices we will have diagrams like:
\begin{align}
\mathrm{Tr} (\rho_A(q) \sigma_A(q)) = 
\,
\tikz[baseline=0ex]{
	\draw[dashed] (0,0.2)-- (0,0);
	\draw[dashed] (0,0) -- (1,0);
	\draw[dashed]  (1,0.2) -- (1,0);
		\draw (-0.3,0.5) circle (0.) node[anchor=center] {\footnotesize$i_{q}$};
	\draw (1.3,0.5) circle (0.) node[anchor=center] {\footnotesize$i_{q}$};
	\draw (0.1,0.5) circle (0.) node[anchor=center] {\footnotesize$ \alpha_{\bar{q}} $};
	\draw (0.9,0.5) circle (0.) node[anchor=center] {\footnotesize$ \alpha_{\bar{q}} $};
	\draw (-0.2,0.2)-- (-0.2,-0.25);
	\draw (1.2,-0.1)-- (1.2,0.2);
	\draw[dashed,BrickRed] (2,0.2)-- (2,0);
	\draw[dashed,BrickRed] (2,0) -- (3,0);
	\draw[dashed,BrickRed]  (3,0.2) -- (3,0);
	\draw[BrickRed] (1.8,0.2)-- (1.8,-0.1);
	\draw[BrickRed] (3.2,-0.25)-- (3.2,0.2);
		\draw (1.7,0.5) circle (0.) node[anchor=center] {\footnotesize$i_{q}$};
	\draw (3.3,0.5) circle (0.) node[anchor=center] {\footnotesize$i_{q}$};
	\draw (2.1,0.5) circle (0.) node[anchor=center] {\footnotesize$ \alpha_{\bar{q}} $};
	\draw (2.9,0.5) circle (0.) node[anchor=center] {\footnotesize$ \alpha_{\bar{q}} $};
	\draw (1.2,-0.1)-- (1.8,-0.1);
	\draw (-0.2,-0.25) -- (3.2,-0.25);
}\ .
\end{align}
 There is only one term for the ensemble averaging process, as the black and red lines are averaged separately.
\begin{align}
\overline{\mathrm{Tr} (\rho_A(q) \sigma_A(q))} &=
\,
\tikz[scale=0.8,baseline=0.5ex]{
	\draw[dashed] (0,0) -- (1,0);
	\draw (-0.2,0.)-- (-0.2,-0.25);
	\draw (1.2,-0.1)-- (1.2,0.);
	\draw (-0.3,0.5) circle (0.) node[anchor=center] {\footnotesize$q$};
	\draw (0.6,0.2) circle (0.) node[anchor=center] {\footnotesize$ \bar{q} $};
	\draw[dashed,BrickRed] (2,0) -- (3,0);
	\draw[BrickRed] (1.8,0.)-- (1.8,-0.1);
	\draw[BrickRed] (3.2,-0.25)-- (3.2,0.);
	\draw (1.2,-0.1)-- (1.8,-0.1);
	\draw (-0.2,-0.25) -- (3.2,-0.25);
	\draw[dashed] (1.0,0) arc (0:180:0.5);
	\draw[dashed,BrickRed] (3.0,0) arc (0:180:0.5);
	\draw (1.7,0.5) circle (0.) node[anchor=center] {\footnotesize$q$};
	\draw (2.5,0.2) circle (0.) node[anchor=center] {\footnotesize$ \bar{q} $};
	\draw (1.2,0) arc (0:180:0.7);
	\draw[BrickRed] (3.2,0) arc (0:180:0.7);
},
\ \,
\label{overlap_diagram}
\end{align}
The above expression gives $d_A^{-1}(q)$. For higher powers of $\sigma_A$, we have the following diagram,
\begin{align}
	\mathrm{Tr}(\rho_A(q) \left(\sigma_A(q)\right)^{n-1})
	=
	\tikz[scale=0.75,baseline=-0.5ex]{
		\draw[dashed] (0,0) -- (1,0);
		\draw (0,0)-- (0,.15);
		\draw (1,0)-- (1,.15);
		\draw (-0.2,0.15)-- (-0.2,-0.35);
		\draw (-0.3,0.5) circle (0.) node[anchor=center] {\footnotesize$q$};
		\draw (0.6,0.2) circle (0.) node[anchor=center] {\footnotesize$ \bar{q} $};
		\draw (1.2,-0.15)-- (1.2,0.15);
		\draw[dashed,BrickRed] (2,0) -- (3,0);
		\draw[BrickRed] (2,0)-- (2,0.15);
		\draw[BrickRed] (3,0)-- (3,0.15);
		\draw[BrickRed] (1.8,-.15)-- (1.8,0.15);
		\draw[BrickRed] (3.2,-0.15)-- (3.2,0.15);
		\draw[dashed,BrickRed] (5,0) -- (6,0);
		\draw[BrickRed] (5,0)-- (5,0.15);
		\draw[BrickRed] (6,0)-- (6,0.15);
		\draw[BrickRed] (4.8,0.15)-- (4.8,-0.15);
		\draw[BrickRed] (6.2,-0.35)-- (6.2,0.15);
		\draw (1.2,-0.15)--(1.8,-0.15);
		\draw[BrickRed] (3.2,-0.15)--(3.45,-0.15);
		\draw[black] (-0.2,-0.35)-- (1.8,-0.35);
		\draw[BrickRed] (1.8,-0.35)-- (6.2,-0.35);
		\draw (1.7,0.5) circle (0.) node[anchor=center] {\footnotesize$q$};
		\draw (2.5,0.2) circle (0.) node[anchor=center] {\footnotesize$ \bar{q} $};
			\draw (6.6,0.5) circle (0.) node[anchor=center] {\footnotesize$q$};
		\draw (5.6,0.5) circle (0.) node[anchor=center] {\footnotesize$ \bar{q} $};
		\draw[BrickRed] (4.5,-0.15)--(4.8,-0.15);
		\node[] at (4.,0.2) {$\cdots$};
	} \, .
\end{align}
Due to the presence of only one copy of $\rho_A$, and the disallowance of contracting the black and red indices during ensemble averaging, we need to contract the first density matrix with itself. Consequently, the summation runs only over a subgroup of permutations that stabilize the first element (black lines) and permute the red lines, denoted as $\mathbbm{1}\times S_{n-1}$. This can be inferred from the following diagram:
 \begin{align}
 	\tikz[scale=0.5,baseline=-0.5ex]{
 		\draw[dashed] (0,0) -- (1,0);
 		\draw (-0.2,0.)-- (-0.2,-0.35);
 		\draw (1.2,-0.15)-- (1.2,0.);
 		\draw (-0.3,0.5) circle (0.) node[anchor=center] {\footnotesize$q$};
 		\draw (0.5,0.1) circle (0.) node[anchor=center] {\footnotesize$ \bar{q} $};
 		\draw[dashed,BrickRed] (2,0) -- (3,0);
 		\draw[BrickRed] (2,0)-- (2,0.15);
 		\draw[BrickRed] (3,0)-- (3,0.15);
 		\draw[BrickRed] (1.8,-.15)-- (1.8,0.15);
 		\draw[BrickRed] (3.2,-0.15)-- (3.2,0.15);
 		\draw[dashed,BrickRed] (5,0) -- (6,0);
 		\draw[BrickRed] (5,0)-- (5,0.15);
 		\draw[BrickRed] (6,0)-- (6,0.15);
 		\draw[BrickRed] (4.8,0.15)-- (4.8,-0.15);
 		\draw[BrickRed] (6.2,-0.35)-- (6.2,0.15);
 		\draw (1.2,-0.15)--(1.8,-0.15);
 		\draw[BrickRed] (3.2,-0.15)--(3.45,-0.15);
 		\draw (-0.2,-0.35)-- (6.2,-0.35);
 		\draw[BrickRed] (4.5,-0.15)--(4.8,-0.15);
 		\node[] at (4.,0.2) {$\cdots$};
 		\draw[dashed] (1.0,0) arc (0:180:0.5);
 		\draw (1.2,0) arc (0:180:0.7);
 		\draw (1.7,0.5) circle (0.) node[anchor=center] {\footnotesize$q$};
 		\draw (2.5,0.2) circle (0.) node[anchor=center] {\footnotesize$ \bar{q} $};
 		\draw (6.6,0.5) circle (0.) node[anchor=center] {\footnotesize$q$};
 		\draw (5.6,0.5) circle (0.) node[anchor=center] {\footnotesize$ \bar{q} $};
 	} 
 	=
 	\tikz[scale=0.5,baseline=-0.5ex]{
 		\draw[dashed] (0.5,0) -- (1.5,0);
 		\draw (0.9,0.1) circle (0.) node[anchor=center] {\footnotesize$ \bar{q} $}; 
 		\draw[dashed,BrickRed] (2,0) -- (3,0);
 		\draw[BrickRed] (2,0)-- (2,0.15);
 		\draw[BrickRed] (3,0)-- (3,0.15);
 		\draw[BrickRed] (1.8,-.35)-- (1.8,0.15);
 		\draw[BrickRed] (3.2,-0.15)-- (3.2,0.15);
 		\draw[dashed,BrickRed] (5,0) -- (6,0);
 		\draw[BrickRed] (5,0)-- (5,0.15);
 		\draw[BrickRed] (6,0)-- (6,0.15);
 		\draw[BrickRed] (4.8,0.15)-- (4.8,-0.15);
 		\draw[BrickRed] (6.2,-0.35)-- (6.2,0.15);
 		\draw[BrickRed] (3.2,-0.15)--(3.45,-0.15);
 		\draw (1.8,-0.35)-- (6.2,-0.35);
 		\draw[BrickRed] (4.5,-0.15)--(4.8,-0.15);
 		\node[] at (4.,0.2) {$\cdots$};
 		\draw[dashed] (1.5,0) arc (0:180:0.5);
 		\draw (1.7,0.5) circle (0.) node[anchor=center] {\footnotesize$q$};
 		\draw (2.5,0.2) circle (0.) node[anchor=center] {\footnotesize$ \bar{q} $};
 		\draw (6.6,0.5) circle (0.) node[anchor=center] {\footnotesize$q$};
 		\draw (5.6,0.5) circle (0.) node[anchor=center] {\footnotesize$ \bar{q} $}
 	} .
 \end{align}
   The ensemble average for higher powers of $\sigma_A$ can be expressed as a sum over the broken permutation group, $\tau \in \mathbbm{1}\times S_{n-1}$. 
 \begin{align}
 	\overline{ \mathrm{Tr} ( \rho_A(q) (\sigma_A(q))^{n-1})}
 	\nonumber
 	&= \frac{1}{(d_A(q) d_{\bar{A}}(\bar{q}))^n}\\
 	&\times\sum_{\tau \in \mathbbm{1}\times S_{n-1}} (d_A(q))^{C(\eta^{-1} \circ \tau)}(d_{\bar{A}}(\bar{q}))^{C(\tau)}.
 	\label{relative_sum}
 \end{align}
 To maximize the exponent, we should maximize the number of loops in the diagrams. The diagrams that maximize the total number of loops are noncrossing ones that only act on the $(n-1)$ red indices. The number of such non-crossing permutations $NC_{n-1}$ with $C(\eta^{-1}\circ \tau) = k$ is given by the Narayana number $N_{n-1,k}$. Therefore, we can rewrite the sum as:
\begin{align}
\overline{\mathrm{Tr}( \rho_A(q) (\sigma_A(q))^{n-1})} &=
\frac{1}{(d_A(q) d_{\bar{A}}(\bar{q}))^n}
\nonumber
\\
&\times
\sum_{k=1}^{n-1} N_{n-1,k}(d_A(q))^{k}(d_{\bar{A}}(\bar{q}))^{n+1-k},
\nonumber
\end{align}
The above expression can  be written as a hypergeometric function
\begin{align}
	\overline{\mathrm{Tr}(\rho_A(q) \left(\sigma_A(q)\right)^{n-1})}
	& \, 
	\label{relative_hyper}=\begin{cases}
		(d_A(q))^{1-n}\, _2F_1\left(1-n,2-n;2;\frac{d_A(q)}{d_{\bar{A}}(\bar{q})}\right)., &d_A(q) < (d_{\bar{A}}(\bar{q})
		\\
		(d_{\bar{A}}(\bar{q}))^{2-{n}}(d_A(q))^{-1} \, _2F_1\left(1-{n},2-{n};2;\frac{d_{\bar{A}}(\bar{q})}{d_A(q)}\right), &d_A(q) > d_{\bar{A}}(\bar{q})
	\end{cases} 
\end{align}
Taking the $n \rightarrow 1$ limit, we find
\begin{align}
	\lim_{n \rightarrow 1}\frac{1}{1-n} \overline{\log\left( \mathrm{Tr}(\rho_A(q) \left(\sigma_A(q)\right)^{n-1})\right) }= \begin{cases}
		\log \left(d_A(q)\right)  +1 +\left(\frac{d_{\bar{A}}(\bar{q})}{d_A(q)}-1 \right)\log\left( 1-\frac{d_A(q)}{d_{\bar{A}}(\bar{q})}\right) , & d_A(q) < d_{\bar{A}}(\bar{q})
		\\
		\infty,& d_{\bar{A}}(\bar{q})<d_A(q) 
	\end{cases}.
\end{align}
Therefore, the ensemble average of
the symmetry-resolved relative entropy obtained by \footnote{In general, the ensemble average and logarithm do not commute, so a further replica trick is needed. However, in the case of large Hilbert space dimensions, these operations approximately commute 
	\cite{Kudler-Flam:2021rpr}.}
\begin{align}
	\overline{D(\rho_A || \sigma_A)} = \begin{cases}
		1+\frac{d_A(q) }{2 d_{\bar{A}}(\bar{q})}+\left(\frac{d_{\bar{A}}(\bar{q})}{d_A(q) }-1\right) \log
		\left( 1-\frac{d_A(q) }{d_{\bar{A}}(\bar{q})}\right) , & d_A(q)  < d_{\bar{A}}(\bar{q})
		\\
		\infty, & d_A(q)  > d_{\bar{A}}(\bar{q})
	\end{cases}.
	\label{relative_random}
\end{align}
 This formula represents the final result of symmetry-resolved relative entropy for Haar random states.

 When the ratio $d_A(q)/d_{\bar{A}}(\bar{q})$ is very small, $d_A(q)/d_{\bar{A}}(\bar{q})\ll 1$, the symmetry-resolved relative entropy can be approximated as $\overline{D(\rho_A || \sigma_A)} \approx d_A(q)/d_{\bar{A}}(\bar{q})$. This approximation suggests that the symmetry-resolved density matrices become almost indistinguishable when we have limited information about the symmetry sector, as we would expect.
 
 The symmetry-resolved relative entropy increases monotonically as $d_A(q)/d_{\bar{A}}(\bar{q})$ increases, and it approaches a value of 3/2 as $d_A(q)$ approaches $d_{\bar{A}}(\bar{q})$. This behavior reflects the monotonicity of symmetry-resolved relative entropy under the partial trace.

It is worth noting that the relative entropy can be infinite if the support of $\rho_A$ intersects with the kernel of  $\sigma_A$, indicating that the two states are orthogonal and hence maximally distinguishable. In other words, the relative entropy is finite, if the support of $\rho_A$ is contained within the support of 
$\sigma_A$. In the symmetry-resolved relative entropy of random states, the Wishart ensemble has a rank of at most $\min(d_A(q), d_{\bar{A}}(\bar{q}))$, and when  $d_A(q) > d_{\bar{A}}(\bar{q})$, every reduced state on \(\mathcal{H}_A(q)\) in the ensemble will have a deficiency of rank with $d_A(q) - d_{\bar{A}}(\bar{q})$ zero eigenvalues. Consequently, it is highly improbable that two independent states, $\rho_A(q)$ and $\sigma_A(q)$, will have the same support. Specifically, the support of $\rho_A(q)$ will not be contained within the support of $\sigma_A(q)$. This is why the symmetry-resolved relative entropy becomes formally infinite in this regime; there exists a measurement that allows us to effortlessly distinguish between $\rho_A(q)$ and $\sigma_A(q)$.

\section{Conclusion}\label{Sec-4}
	
	In this manuscript, we have utilized a large-$N$ diagrammatic technique to calculate symmetry-resolved
	relative entropies. These entropies provide a measure of distinguishability within a given symmetry sector. We have focused on  U(1) symmetric random states drawn from the Wishart ensemble. Using the replica trick formalism, we have obtained the symmetry-resolved Page's curve and symmetry-resolved relative entropies. These results deepen our understanding of the properties of these random states.

	Our results have shown that within each symmetry sector, in the Abelian case, the symmetry-resolved relative entropy monotonically
	increases with $d_A/d_B$ and reaches a value of  $3/2$
	when $d_A=d_B$. This behavior demonstrates
	the monotonicity of symmetry-resolved relative entropy under the partial trace. Moreover, we found that when $d_A(q) > d_{\bar{A}}(\bar{q})$, the symmetry-resolved relative entropy becomes formally infinite. 
	
	It would be interesting to generalize  our formalism to non-abelian random states \cite{Bianchi:2024aim}, the asymmetric case \cite{Ares:2022koq} for random states \cite{Ares:2023ggj}  or compute the
	\textit{Symmetry-resolved sandwiched R\'enyi relative entropies}\cite{Ghasemi:2024dim} for random states. We will go back to these problems in the near future.


	\subsubsection*{Acknowledgments}

	It is my pleasure to thank Pasquale Calabrese,  Sepideh Mohammadi, Ahmad Moradpouri, and Sara Murciano for reading our manuscript and providing insightful comments.



\begin{thebibliography}{}
	\bibitem{Mehta:2004RMT}
	M.~L. Mehta, \href{http://dx.doi.org/10.1016/S0079-8169(04)80088-6}{{\em
			{Random Matrices}}}, vol.~142 of {\em Pure and Applied Mathematics}.
	\newblock Academic Press, 3rd~ed., 2004.
	
	
	\bibitem{Eynard:2015aea}
	B.~Eynard, T.~Kimura and S.~Ribault,
	Random matrices,
	[arXiv:1510.04430 [math-ph]].
	
	
	
	
	\bibitem{DAlessio:2015qtq}
	L.~D'Alessio, Y.~Kafri, A.~Polkovnikov and M.~Rigol,
	From quantum chaos and eigenstate thermalization to statistical mechanics and thermodynamics,
	Adv. Phys. \textbf{65}, no.3, 239-362 (2016)
	
	\bibitem{Collins:2015qtqI}
	B.~Collins, I.~Nechita, 
	Random matrix techniques in quantum information theory,
	J. Math. Phys. \textbf{ 57}, 015215 (2016)
	

\bibitem{Popescu:2006gts}
S.~ Popescu, A.J.~Short, A.~Winter
The foundations of statistical mechanics from entanglement: Individual states vs. averages,
Nat. Phys. \textbf{2}, 754 (2006)


\bibitem{Gogolin:2015gts}
C.~Gogolin and J.~Eisert,
Equilibration, thermalisation, and the emergence of statistical mechanics in closed quantum systems,
Rept. Prog. Phys. \textbf{79}, no.5, 056001 (2016)



	\bibitem{Page:1993wv}
D.~N.~Page,
Information in black hole radiation,
Phys. Rev. Lett. \textbf{71}, 3743-3746 (1993)



\bibitem{Hayden:2007cs}
P.~Hayden and J.~Preskill,
Black holes as mirrors: Quantum information in random subsystems,
JHEP \textbf{09}, 120 (2007)


\bibitem{Page:1993df}
D.~N.~Page,
Average entropy of a subsystem,
Phys. Rev. Lett. \textbf{71}, 1291-1294 (1993)


\bibitem{Bianchi:2021aui}
E.~Bianchi, L.~Hackl, M.~Kieburg, M.~Rigol and L.~Vidmar,
Volume-Law Entanglement Entropy of Typical Pure Quantum States,
PRX Quantum \textbf{3}, no.3, 030201 (2022)



\bibitem{Vedral:2002zz}
V.~Vedral,
The role of relative entropy in quantum information theory,
Rev. Mod. Phys. \textbf{74}, 197-234 (2002)

\bibitem{Wehrl:1978zz}
A.~ Wehrl,
General properties of entropy,
Rev. Mod. Phys. \textbf{50}, 221 – Published 1 April (1978)



\bibitem{Faulkner:2016mzt}
T.~Faulkner, R.~G.~Leigh, O.~Parrikar and H.~Wang,
JHEP \textbf{09}, 038 (2016)


\bibitem{Balakrishnan:2017bjg}
S.~Balakrishnan, T.~Faulkner, Z.~U.~Khandker and H.~Wang,
A General Proof of the Quantum Null Energy Condition,
JHEP \textbf{09}, 020 (2019)

\bibitem{Casini:2016udt}
H.~Casini, E.~Teste and G.~Torroba,
Relative entropy and the RG flow,
JHEP \textbf{03}, 089 (2017)


\bibitem{Casini:2019qst}
H.~Casini, S.~Grillo and D.~Pontello,
Relative entropy for coherent states from Araki formula,
Phys. Rev. D \textbf{99}, no.12, 125020 (2019)









\bibitem{Lashkari:2014yva}
N.~Lashkari,
Relative Entropies in Conformal Field Theory,
Phys. Rev. Lett. \textbf{113}, 051602 (2014)

\bibitem{Lashkari:2015dia}
N.~Lashkari,
Modular Hamiltonian for Excited States in Conformal Field Theory,
Phys. Rev. Lett. \textbf{117}, no.4, 041601 (2016)

\bibitem{Ruggiero:2016khg}
P.~Ruggiero and P.~Calabrese,
Relative Entanglement Entropies in 1+1-dimensional conformal field theories,
JHEP \textbf{02}, 039 (2017)


	\bibitem{Sarosi:2016oks}
		G.~S\'arosi and T.~Ugajin,
Relative entropy of excited states in two dimensional conformal field theories,''
		JHEP \textbf{07}, 114 (2016)


\bibitem{Ghasemi:2024wcq}
M.~Ghasemi,
Left-Right Relative Entropy,
[arXiv:2411.09406 [hep-th]].


\bibitem{Blanco:2013joa}
D.~D.~Blanco, H.~Casini, L.~Y.~Hung and R.~C.~Myers,
Relative Entropy and Holography,
JHEP \textbf{08}, 060 (2013)


 \bibitem{Jafferis:2014lza}
D.~L.~Jafferis and S.~J.~Suh,
The Gravity Duals of Modular Hamiltonians,
 JHEP \textbf{09}, 068 (2016)


\bibitem{Jafferis:2015del}
D.~L.~Jafferis, A.~Lewkowycz, J.~Maldacena and S.~J.~Suh,
Relative entropy equals bulk relative entropy,



\bibitem{Lashkari:2016idm}
 N.~Lashkari, J.~Lin, H.~Ooguri, B.~Stoica and M.~Van Raamsdonk,
Gravitational positive energy theorems from information inequalities,
 PTEP \textbf{2016}, no.12, 12C109 (2016)





\bibitem{Casini:2008cr}
H.~Casini,
Relative entropy and the Bekenstein bound,
Class. Quant. Grav. \textbf{25}, 205021 (2008)

\bibitem{Wall:2011hj}
A.~C.~Wall,
A proof of the generalized second law for rapidly changing fields and arbitrary horizon slices,
Phys. Rev. D \textbf{85}, 104049 (2012)


	\bibitem{Bousso:2014sda}
	R.~Bousso, H.~Casini, Z.~Fisher and J.~Maldacena,
Proof of a Quantum Bousso Bound,
	Phys. Rev. D \textbf{90}, no.4, 044002 (2014)


\bibitem{Bousso:2014uxa}
	R.~Bousso, H.~Casini, Z.~Fisher and J.~Maldacena,
Entropy on a null surface for interacting quantum field theories and the Bousso bound,
	Phys. Rev. D \textbf{91}, no.8, 084030 (2015)




\bibitem{Longo:2018zib}
R.~Longo and F.~Xu,
Comment on the Bekenstein bound,
J. Geom. Phys. \textbf{130}, 113-120 (2018)


\bibitem{Kudler-Flam:2021rpr}
J.~Kudler-Flam,
Relative Entropy of Random States and Black Holes,
Phys. Rev. Lett. \textbf{126}, no.17, 171603 (2021)


\bibitem{Kudler-Flam:2021alo}
J.~Kudler-Flam, V.~Narovlansky and S.~Ryu,
Distinguishing Random and Black Hole Microstates,
PRX Quantum \textbf{2}, no.4, 040340 (2021)

\bibitem{Shapourian:2020mkc}
H.~Shapourian, S.~Liu, J.~Kudler-Flam and A.~Vishwanath,
Entanglement Negativity Spectrum of Random Mixed States: A Diagrammatic Approach,
PRXQuantum \textbf{2}, no.3, 030347 (2021)


\bibitem{Göran:1974rpr}
G.~ Lindblad ,
Completely Positive Maps
and Entropy Inequalities,
Commun. math. Phys.\textbf{40}, 147 (1975)


\bibitem{Goldstein:2017bua}
M.~Goldstein and E.~Sela,
Symmetry-resolved entanglement in many-body systems,
Phys. Rev. Lett. \textbf{120}, no.20, 200602 (2018)

\bibitem{lukin:2018}
A. Lukin, M. Rispoli, R. Schittko, M. E. Tai, A. M. Kaufman, S. Choi, V. Khemani, J. Leonard, and M. Greiner,
{\it Probing entanglement in a many-body localized system},
\href{https://science.sciencemag.org/content/364/6437/256/tab-figures-data}{Science {\bf 364}, 6437 (2019)}.


\bibitem{wv-03}
H. M. Wiseman and J. A. Vaccaro, {\it Entanglement of Indistinguishable Particles Shared between Two Parties}, 
\href{https://doi.org/10.1103/PhysRevLett.91.097902}{Phys. Rev. Lett. {\bf 91}, 097902 (2003)}.

\bibitem{bhd-18}
H. Barghathi, C. M. Herdman, and A. Del Maestro, {\it R\'enyi Generalization of the Accessible Entanglement Entropy}, 
\href{https://doi.org/10.1103/PhysRevLett.121.150501}{Phys. Rev. Lett. {\bf 121}, 150501 (2018)};\\

\bibitem{kusf2}
M. Kiefer-Emmanouilidis, R. Unanyan, J. Sirker, and M. Fleischhauer, {\it Evidence for unbounded growth of the number entropy in many-body localized phases},
\href{https://dx.doi.org/10.1103/PhysRevLett.124.243601}{Phys. Rev. Lett. {\bf 124}, 243601 (2020)}.



\bibitem{Xavier:2018kqb}
J.~C.~Xavier, F.~C.~Alcaraz and G.~Sierra,
Equipartition of the entanglement entropy,
Phys. Rev. B \textbf{98}, no.4, 041106 (2018)

\bibitem{Capizzi:2020jed}
L.~Capizzi, P.~Ruggiero and P.~Calabrese,
Symmetry resolved entanglement entropy of excited states in a CFT,
J. Stat. Mech. \textbf{2007}, 073101 (2020)


\bibitem{Calabrese:2021wvi}
P.~Calabrese, J.~Dubail and S.~Murciano,
Symmetry-resolved entanglement entropy in Wess-Zumino-Witten models,
JHEP \textbf{10}, 067 (2021)


\bibitem{Ghasemi:2022jxg}
M.~Ghasemi,
Universal thermal corrections to symmetry-resolved entanglement entropy and full counting statistics,
JHEP \textbf{05}, 209 (2023)

\bibitem{Ares:2022gjb}
F.~Ares, P.~Calabrese, G.~Di Giulio and S.~Murciano,
Multi-charged moments of two intervals in conformal field theory,
JHEP \textbf{09}, 051 (2022)

\bibitem{DiGiulio:2022jjd}
G.~Di Giulio, R.~Meyer, C.~Northe, H.~Scheppach and S.~Zhao,
On the boundary conformal field theory approach to symmetry-resolved entanglement,
SciPost Phys. Core \textbf{6}, 049 (2023)

\bibitem{Kusuki:2023bsp}
Y.~Kusuki, S.~Murciano, H.~Ooguri and S.~Pal,
Symmetry-resolved entanglement entropy, spectra \& boundary conformal field theory,
JHEP \textbf{11}, 216 (2023)

\bibitem{Gaur:2023yru}
H.~Gaur and U.~A.~Yajnik,
Multi-charged moments and symmetry-resolved R\'enyi entropy of free compact boson for multiple disjoint intervals,
JHEP \textbf{01}, 042 (2024)








\bibitem{Laflorencie2014}
N.~Laflorencie and S.~Rachel, 
\newblock {\it Spin-resolved entanglement spectroscopy of critical spin chains and Luttinger liquids},
\newblock \href{http://dx.doi.org/10.1088/1742-5468/2014/11/P11013}{J. Stat. Mech. P11013 (2014)}.



\bibitem{Belin:2013uta}
A.~Belin, L.~Y.~Hung, A.~Maloney, S.~Matsuura, R.~C.~Myers and T.~Sierens,
Holographic Charged Renyi Entropies,
JHEP \textbf{12}, 059 (2013)

\bibitem{Zhao:2020qmn}
S.~Zhao, C.~Northe and R.~Meyer,
Symmetry-resolved entanglement in AdS$_{3}$/CFT$_{2}$ coupled to U(1) Chern-Simons theory,
JHEP \textbf{07}, 030 (2021)

\bibitem{Weisenberger:2021eby}
K.~Weisenberger, S.~Zhao, C.~Northe and R.~Meyer,
Symmetry-resolved entanglement for excited states and two entangling intervals in AdS$_{3}$/CFT$_{2}$,
JHEP \textbf{12}, 104 (2021)


\bibitem{Zhao:2022wnp}
S.~Zhao, C.~Northe, K.~Weisenberger and R.~Meyer,
`Charged moments in W$_{3}$ higher spin holography,
JHEP \textbf{05}, 166 (2022)




\bibitem{DiGiulio:2023nvz}
G.~Di Giulio and J.~Erdmenger,
Symmetry-resolved modular correlation functions in free fermionic theories,
JHEP \textbf{07}, 058 (2023)
%

\bibitem{Milekhin:2021lmq}
A.~Milekhin and A.~Tajdini,
Charge fluctuation entropy of Hawking radiation: A replica-free way to find large entropy,
SciPost Phys. \textbf{14}, no.6, 172 (2023)


\bibitem{Hejazi:2021yhz}
K.~Hejazi and H.~Shapourian,
Symmetry-protected entanglement in random mixed states,
Phys. Rev. A \textbf{106}, no.5, 052428 (2022)


\bibitem{Horvath:2023xoh}
D.~X.~Horvath, S.~Fraenkel, S.~Scopa and C.~Rylands,
Charge-resolved entanglement in the presence of topological defects,
Phys. Rev. B \textbf{108}, no.16, 165406 (2023)

\bibitem{Fossati:2023zyz}
M.~Fossati, F.~Ares and P.~Calabrese,
Symmetry-resolved entanglement in critical non-Hermitian systems,
Phys. Rev. B \textbf{107}, no.20, 205153 (2023)

\bibitem{Feldman:2024qif}
N.~Feldman, J.~Knaute, E.~Zohar and M.~Goldstein,
Superselection-resolved entanglement in lattice gauge theories: a tensor network approach,
JHEP \textbf{05}, 083 (2024)









\bibitem{Chen:2021pls}
H.~H.~Chen,
Symmetry decomposition of relative entropies in conformal field theory,
JHEP \textbf{07}, 084 (2021)

\bibitem{Capizzi:2021zga}
L.~Capizzi and P.~Calabrese,
Symmetry resolved relative entropies and distances in conformal field theory,
JHEP \textbf{10} (2021), 195

\bibitem{KREWERAS1972333}
G.~ Kreweras,
Sur les partitions non
croisees d'un cycle,
Discrete Mathematics , volume 1 , number 4 , pages 333 – 350 , 1972 


\bibitem{SIMION2000367}
R.~Simion,
 Noncrossing partitions,
Discrete Mathematics, volume 217, numbers 1–3, pages 367–409, April 2000.



\bibitem{Bianchi:2019stn}
E.~Bianchi and P.~Dona,
Typical entanglement entropy in the presence of a center: Page curve and its variance,
Phys. Rev. D \textbf{100}, no.10, 105010 (2019)



\bibitem{Murciano:2022lsw}
S.~Murciano, P.~Calabrese and L.~Piroli,
Symmetry-resolved Page curves,
Phys. Rev. D \textbf{106}, no.4, 046015 (2022)

\bibitem{Lau:2022hvc}
P.~H.~C.~Lau, T.~Noumi, Y.~Takii and K.~Tamaoka,
Page curve and symmetries,
JHEP \textbf{10}, 015 (2022)


\bibitem{Li:2023zgy}
P.~Li and Y.~Ling,
Refined symmetry-resolved Page curve and charged black holes,
Chin. Phys. C \textbf{48}, no.5, 053109 (2024)


\bibitem{Bianchi:2024aim}
E.~Bianchi, P.~Dona and R.~Kumar,
Non-abelian symmetry-resolved entanglement entropy,
[arXiv:2405.00597 [quant-ph]].


\bibitem{Ares:2022koq}
F.~Ares, S.~Murciano and P.~Calabrese,
Entanglement asymmetry as a probe of symmetry breaking,
Nature Commun. \textbf{14}, no.1, 2036 (2023)

\bibitem{Ares:2023ggj}
F.~Ares, S.~Murciano, L.~Piroli and P.~Calabrese,
Entanglement asymmetry study of black hole radiation,
Phys. Rev. D \textbf{110}, no.6, L061901 (2024)

\bibitem{Ghasemi:2024dim}
M.~Ghasemi, 
Symmetry-resolved sandwiched R\'enyi relative entropies of random states (In progress)








\end{thebibliography}
\end{document}